\documentclass[event]{sigirforum}
\usepackage{subcaption}
\usepackage{caption}
\usepackage{graphics}

\def\pubissue{Vol. 60 No. 1 -- June 2026}

\def\eventdate{26 March 2026}
\def\eventurl{https://sites.google.com/view/chiir-gaias-workshop/home}

\begin{document}
\title{Report on CHIIR 2026 Workshop on Generative AI and Academic Search (GAI\&AS)}

\authors{
\author[yiifan@student.ubc.ca]{Yifan Liu}{University of British Columbia}{Canada}
\and
\author[jarguell@email.unc.edu]{Jaime Arguello}{University of North Carolina at Chapel Hill}{USA}
\and
\author[orland.hoeber@uregina.ca]{Orland Hoeber}{University of Regina}{Canada}
\and
\author[imliuc@pku.edu.cn]{Chang Liu}{Peking University}{China}
\and
\author[rieh@ischool.utexas.edu]{Soo Young Rieh}{University of Texas at Austin}{USA}
\and
\author[luanne.sinnamon@ubc.ca]{Luanne Sinnamon}{University of British Columbia}{Canada}
\and
\authorothers{
Dean Alvarez,
Susan Archambault,
Rob Capra, 
Henson Chen,
Charles Costa,
Anita Crescenzi,
Zhitong (Klara) Guan,
Jacek Gwizdka,
Pao-Pei Huang,
Gavindya Jayawardena,
Ghazal Kalhor,
Dagmar Kern,
Oliver Koop,
Alice Li,
Afra Mashhadi,
Gaohui Meng,
Marta Micheli,
Anil B. Murthy,
Kevin Schott,
Sebastian Schultheiß,
Jiwoo Seo,
Phaneendra Sivangula,
Frans van der Sluis,
Xiaoxuan Song,
Silang Wang,
Dan Zhang}
}

\maketitle 
\begin{abstract}
This report summarizes the CHIIR 2026 Workshop on Generative AI and Academic Search (GAI\&AS), which examined how GenAI is reshaping academic search systems and research practices. The workshop brought together researchers in human information interaction and information retrieval to explore key challenges and opportunities in designing and evaluating future academic search systems that integrate GenAI, moving beyond traditional document retrieval to support summarization, recommendation, synthesis, and conversational interaction. Participants' interests and discussions focused on three thematic clusters: foundations and principles, applications and opportunities, and search-as-learning. Across these themes, the workshop highlighted the importance of academic search systems in supporting transparency, credibility, research integrity, and long-term scholarly needs, as well as in fostering higher-order cognitive processes. Participants discussed guiding theories, design principles, methodological approaches, partnerships, and community-building efforts aimed at advancing human-centered GenAI-enhanced academic search systems. Overall, the workshop demonstrated strong community interest and a diverse range of ongoing and emerging research initiatives at the intersection of GenAI and academic search. 
\end{abstract}

\section{Introduction}
Recent developments in GenAI are reshaping the landscape of academic search. Systems that once focused primarily on document retrieval now increasingly support summarization, recommendation, synthesis, and conversational interaction, thereby extending their role in research and learning workflows~\citep{adams_artificial_2022, heidt_artificial-intelligence_2023, wagner_artificial_2022}. As these capabilities become embedded in academic search engines, bibliographic databases, and digital library environments, they unsettle a basic assumption that has long organized work in information retrieval and academic search: that the search system helps users locate relevant sources, while the interpretive and epistemic work of evaluating, connecting, and using those sources remains more visibly in human hands.

That shift matters particularly in academic search. In contrast to many everyday search scenarios, scholarly work depends on traceability, credibility, and the careful interpretation of evidence~\citep{gusenbauerWhatEveryResearcher2021}. AI-enhanced academic search tools may reduce effort and support discovery in valuable ways~\citep{chavula_searchidea_2023,khalifa_using_2024}, but they also reconfigure the boundaries between retrieval, interpretation, and reasoning~\citep{bolanosArtificialIntelligenceLiterature2024, yang_searchchat_2025, mitra2024sociotechnicalimplicationsgenerativeartificial}. This makes academic search a productive site for examining longstanding concerns in human-centered IR, including learning, cognitive effort, agency, and the design of interfaces that support critical engagement rather than unreflective acceptance. The workshop was motivated by the view that these developments should not be treated merely as a new application area for GenAI, but as an opportunity to revisit assumptions about how search systems support knowledge work.

The CHIIR'26 Workshop on Generative AI and Academic Search (GAI\&AS)~\citep{liuWorkshopGenerativeAI2026} was organized around this emerging tension. Positioned at the intersection of IIR, search-as-learning, information behaviour, and scholarly communication, the workshop used academic search as a lens for reconsidering what search systems are now expected to do and how they should be studied. The sections that follow outline the main topics around which the workshop was organized, before turning to the lightning talks, the thematic clustering activity, and the workshop's discussions of future challenges and opportunities.

\section{Workshop Topics}

The workshop was organized to develop a shared intellectual agenda for research on GenAI in academic search. It approached AI-enhanced academic search as a socio-technical and human-centered research problem involving changing patterns of information seeking, new forms of support for research and learning, and broader epistemic and ethical questions about the division of labor between system and user.

The workshop was organized around four interrelated themes in the call for submissions: supporting research and education; design, evaluation, and human-centered interaction; search-as-learning in the era of GenAI; and search integrity and ethics. Together, these themes addressed the roles GenAI-enhanced academic search systems are beginning to play in ideation, literature review, evidence synthesis, and teaching; the design and evaluation challenges raised when such systems do more than retrieve documents; the ways GenAI may reshape the cognitive, strategic, and metacognitive dimensions of academic search; and the risks associated with fairness, accountability, transparency, ethics (FATE), and cognitive offloading. These themes provided the conceptual structure for the workshop's presentations and discussions, informing the lightning talks, thematic clustering activities, and later reflections on challenges and future directions.

\subsection{Lightning Talks}


\subsubsection{Optimizing Generative AI-Powered Search Systems to Support Human-AI Symbiosis in Creative Academic Search}


Chang Liu, in collaboration with Gaohui Meng and Xiaoxuan Song, described how in creative academic search, users offload low-level tasks to GenAI but their usage drops during insight generation due to low confidence, revealing a mismatch between GenAI capabilities and high-level cognitive needs. In her lightning talk, Dr. Chang Liu introduced two ongoing projects on how to inspire GenAI functionality by revealing user behavior patterns. First, using the Information Problem Space (IPS) framework~\citep{10.1145/3624918.3625326}, Meng and Liu modeled how users decompose, formalize, and formulate information needs into prompts through three layers. Second, Song and Liu examined and identified three types of metacognitive laziness during academic search behaviors: plan-deficit, monitoring-deficit, and regulation-deficit. Two design challenges emerge. First, GenAI should evolve from fact-retrievers to problem-solving scaffolds that help users break down ill-structured problems into manageable sub-problems at the strategy layer. Second, systems should embed metacognitive scaffolds, such as prompting sub-question formulation, guiding output comparison, or encouraging reflection on AI-user knowledge discrepancies, to mitigate each type of laziness. Addressing these challenges can foster adaptive collaboration and support human-AI symbiosis in creative academic search.

\subsubsection {When Answers Are Not Enough: Balancing Generative AI and Human Intelligence in Academic Digital Library Search}


Orland Hoeber's lightning talk started with an overview of how in recent years, significant advances have been made with large language models (LLMs), generative artificial intelligence (GenAI), and retrieval augmented generation (RAG)~\citep{huang-huang-2026}. These technologies have changed the nature of information retrieval from providing ranked lists of resources for searchers to evaluate, to automating relevance assessment and synthesis to produce answers to explicit or implied questions. While such approaches work well when academic searchers have clear end goals (e.g., information lookup tasks, well-defined learning tasks), not all academic searching can be resolved with AI-generated answers.

Complex academic search tasks (those that are ambiguous, multi-faceted, or open ended) are better suited to exploratory search processes where the searcher takes an active role in the search activities~\citep{white-roth-2009}. In such cases, the value is in the learning that occurs during the search journey. The question then is: How can we leverage the power of GenAI to enhance this journey, rather than replace it? 

Hoeber summarized recent work that provides an alternative to RAG for academic search, striking a balance between GenAI automation and human intelligence. He established multi-workspace environments as an important foundation~\citep{gomes_study_2022}, and showed how GenAI could be used to suggest narrow and broad queries based on recent exploratory search activities~\citep{khakshoor-hoeber-2025} and to summarize what has been found in support of cross-session exploratory search~\citep{orin-hoeber-2025,momeni-hoeber-2026}. He advocated for maintaining human agency within academic search processes.

\subsubsection{Remembering Unequally: Global and Disciplinary Bias in LLM Reconstruction of Scholarly Coauthor Lists}


Afra Mashhadi, working with Ghazal Kalhor, reported on an investigation of 
whether large language models (LLMs) memorize scholarly coauthor lists equitably across academic disciplines and world regions. Using a novel set-based metric---Discoverable Name Extraction (DNE)---they evaluate three models of varying scale (DeepSeek R1, Llama 4 Scout, and Mixtral 8$\times$7B) against bibliographic ground truth from OpenAlex and Google Scholar, spanning 1,596 seed authors across 10 disciplines and 8 global regions. Their central finding is unambiguous: highly cited researchers are reconstructed by LLMs at approximately twice the rate of lower-cited peers, a disparity that grows with model size and reflects the disproportionate repetition of prominent names in citation-heavy training corpora. This ``rich get richer'' dynamic risks entrenching existing inequalities in AI-mediated scholarly discovery. Notably, the bias is not universal---Clinical Medicine and parts of Sub-Saharan and North Africa exhibit near-equitable reconstruction, suggesting that more balanced training data can mitigate representational disparities. The paper makes a compelling case that LLM-generated relational outputs must be critically audited before integration into scientometric tools, recommendation systems, or literature review pipelines, where unchecked memorization bias could systematically obscure the work of underrepresented scholars.

\subsubsection{Information Literacy for GenAI Academic Search}

Jiwoo Seo's lightning talk reflected the transition to GenAI-mediated information seeking. During their tenure as an AI Specialist Librarian at Chung-Ang University, they co-developed a 15-week digital literacy curriculum based on a survey of 2,076 students and 101 librarians~\citep{ART003218199}. This curriculum addressed a critical gap by integrating GenAI tool usage with research ethics and critical thinking skills.

Through this work, Seo observed a distinct ``trust gap'': students often over-trusted GenAI, while faculty distrusted it. Because GenAI lowers the barrier to entry for academic search but raises the barrier for verification, Seo's goal is to design educational frameworks that empower users to act as ``critical reviewers'' rather than passive consumers.

To advance GenAI-powered academic search, Seo proposed three key challenges:
\begin{enumerate}[label=(\arabic*)]
\item  Bridging the Gap between AI Capabilities and Academic Integrity: We must address the perception gap regarding AI’s capabilities through robust AI literacy education to ensure appropriate use and honest disclosure in academic writing.
\item Moving beyond One-Off Training to Continuous Literacy: Single-session workshops are insufficient and can leave users with a misunderstood foundation. We need sustainable educational frameworks that adapt to rapid AI advancements.
\item Preserving Human Agency and Competence in Interface Design~\citep{10.1145/3498366.3505816}: Current GenAI tools prioritize speed over intervention. We must design interfaces that keep humans in the loop, preserving their autonomy, competence, and relatedness in academic work.
\end{enumerate}

\subsubsection{Evaluating AI Overviews in a Domain-specific Academic Search Engine}

Kevin Schott's current work examines AI overviews embedded in academic search engine result pages to support early-stage information triage (i.e., the rapid assessment and organization of retrieved materials to serve the task at hand)~\citep{Marshall1997}. Unlike general web search, academic search is fundamentally source-oriented: users seek publications and datasets directly. This makes abstractive summaries potentially useful for triage, but also risky when they omit, distort, or overstate information. In our recent work, we evaluated single SERP-level summaries synthesized from the five top-ranked results in a social science academic search engine. We combined an accuracy assessment across expert-sourced queries with a within-subjects user study comparing interfaces with and without the summary. Results suggested that summaries may reduce mental demand and frustration for some users, while participants descriptively performed slightly fewer clicks and query reformulations. At the same time, the overall effects were mixed. Participants’ feedback highlighted that usefulness depends on context, task, and user characteristics, especially trust in the summary and perceived redundancy. Interpreted through Information Foraging Theory, we view these summaries as possible ``scent concentrators'' that condense cues from several results and may support early-stage triage by reducing the cognitive cost of entering the information patch~\citep{Pirolli2005}. This line of work matters because generative overviews are becoming a common interface layer in academic search engines, yet their benefits and scholarly risks remain insufficiently understood. The community should address two challenges: ensuring reliability and scholarly integrity in generative outputs, and identifying when, for whom, and for which search tasks such features genuinely help.

\subsubsection{Designing for Discernment: Productive Friction in AI-Mediated Academic Search}


As Head of Reference and Instruction at Loyola Marymount University and guest faculty at the University of Washington iSchool, Susan Archambault's work spans library instruction practice and information science education. She oversees information literacy instruction reaching approximately 1,700 first-year students annually.

Archambault argued that AI research tools eliminating uncertainty, exploration, and reflective processes undermine intellectual virtues essential for critical thinking. Her current research maps intellectual virtues, such as curiosity, thoroughness, and intellectual humility, onto the ACRL Framework for Information Literacy, proposing design principles for human-centered AI that preserve decision points where users must practice discernment. Prior empirical work reveals a troubling confidence-knowledge gap: while students' perceived understanding of AI tools increased significantly after instruction, actual knowledge improved minimally~\citep{archambault-etal-2025}. As AI research tools become essential infrastructure, pre-existing equity gaps for transfer and first-generation students risk widening.

A central challenge driving my research agenda is how to design AI research tools that support learning and discernment, not just efficiency, and do so equitably across the full range of users, not just those who are already information-literate.

\subsubsection{Augmenting Sensemaking and Collaboration in AI Academic Search}


Zhitong (Klara) Guan and Soo Young Rieh explained how academic search plays a central role in sensemaking and knowledge construction, through which understanding is gradually built over time. With the rise of GenAI-powered systems, search interactions increasingly include AI-generated summaries, syntheses, thematic groupings, and conversational responses. As AI contributes interpretation rather than simple retrieval, the boundaries between human and machine sensemaking become blurred.

In academic search tasks, GenAI can reduce visibility into how sources are evaluated for relevance, credibility, and usefulness, how themes emerge, how conflicting evidence is handled, and what is excluded. This opacity complicates the evaluation of AI-mediated search using traditional methods, increases the risk of unnoticed hallucinations, and undermines research integrity. It also raises concerns about learning, ownership, and the extent to which original insight can emerge when interpretive work is outsourced. Researchers may rely on AI-generated interpretations without fully understanding or justifying them, limiting opportunities for critically engaged with evidence, reflection, and intellectual ownership.

These challenges are intensified in collaborative research settings. Scholarly collaboration depends on shared norms of scholarly practices such as careful reading, responsible synthesis, and credible interpretation. GenAI introduces a new form of collaboration as researchers work not only with human collaborators but also with AI systems as interpretive labor. When AI-generated syntheses circulate within teams, the basis for trust becomes ambiguous, potentially undermining both efficiency and trust in collaborative knowledge construction. To sustain rigorous scholarship, GenAI systems must be designed and evaluated not only for output quality, but for how they support sensemaking and knowledge construction across individuals and teams.

\subsubsection{How can GenAI Systems Support (and not Hinder) Learning?}

Jaime Arguello argued that future research should ask: Compared to document retrieval systems, do GenAI systems improve self-directed learning?  On one hand, GenAI systems can address highly specific needs.  However, users may be also be tempted to offload processes that may be critical for learning. \cite{urgo_effects_2026} investigated how people use a GenAI system (like ChatGPT) to learn about a complex topic (i.e., diffusion and osmosis). A qualitative analysis of prompts issued to the system revealed an interesting trend.  Participants issued prompts to address needs like those they might address by querying a web search engine. Participants asked for definitions, examples, and explanations.  Interestingly, participants also issued prompts to address needs that GenAI (vs.~document retrieval) systems are uniquely well-suited for.  Participants asked the GenAI to clarify a specific point of confusion (e.g., Are cells permeable or not?); verify a hypothesis (e.g., So osmosis is a type of diffusion?); and test their understanding (e.g., Give me a multiple-choice test.)  Finally, participants asked the GenAI to perform tasks that may have been important for them to do themselves. Participants asked the GenAI system to differentiate between concepts (e.g., Differences between diffusion and osmosis.); provide ideas about what they should learn (e.g., What is important to know?); and simplify a previous response (e.g., Too long, one paragraph.)  Therefore, it remains to be seen whether GenAI systems improve learning.  Future research may need to investigate scaffolding tools or guardrails that encourage users to engage in (vs.~delegate) important processes that improve self-directed learning.

\subsubsection{Agentic Information Seeking}

 
Rob Capra's lightning talk explained how large-language models are opening new possibilities for increasing interactivity and collaboration as part of users’ information seeking process. For example, recent work has explored how users make use of combined chat + search systems, differences in how users interact with AI chat systems compared to web search engines, and users’ trust and verification strategies when using AI chat. While this research provides important insights, it has largely focused on chat-based AI systems that involve one-step-at-a-time (i.e., turn taking) interactions with limited context and limited integration of tools and information external to the LLM.
 
In the past year, tremendous progress has been made in Agentic AI systems that are able to engage in multi-step planning, execution, and external tool use.  Agentic AI systems often run an “agentic loop” that involves gathering context, taking action, verifying results, and getting human-in-the-loop feedback.  Agentic AI is opening new possibilities to support users’ information seeking needs in a more collaborative, iterative way closer to what a human expert could provide.
 
The interactive information retrieval research community needs to do more to explore how agentic AI systems can be developed to support users engaged in complex information seeking and learning tasks. Librarians and information professionals have extensive knowledge and experience that can be leveraged in developing agentic AI systems that effectively help users with their information needs.
\subsubsection {When Search Systems Do More Than Retrieve: Researchers and AI-Enhanced Academic Search}

In Yifan Liu's lightning talk, she introduced an ongoing dissertation project that examines researchers’ interactions with AI-enhanced academic search systems. The talk positioned these systems as a useful lens for understanding how AI is reshaping scholarly information behavior and research practices. As academic search increasingly moves beyond traditional keyword-based retrieval to include semantic search, generated summaries, citation-context exploration, recommendations, and other forms of AI-mediated assistance, search systems are taking on a more active role in the research process~\citep{liuNavigatingEvolvingLandscape2026}. Rather than simply retrieving documents, they now participate in filtering, synthesizing, and structuring scholarly information in ways that may influence how researchers frame problems, explore literature, and make judgments~\citep{liuAITransparencyAcademic2024}.

Liu argued that these developments raise broader questions about whether established information-seeking models remain adequate in environments where search systems do more than retrieve documents. While AI-enhanced systems may improve efficiency and lower barriers to access, they may also redistribute effort, interpretation, and evaluative judgment in ways that are not yet well understood. The talk emphasized that empirical insight remains limited regarding how researchers actually use these systems in authentic research contexts, particularly across different disciplinary backgrounds and levels of search expertise~\citep{liuNavigatingEvolvingLandscape2026}. It also highlighted the importance of examining not only whether researchers adopt these tools, but how they integrate them into real workflows, how they move across multiple systems, and how AI-mediated support may reshape researcher agency, trust, and search strategy.

\subsection{Thematic Clusters}



\begin{figure}[h]
\captionsetup[subfigure]{justification=centering}
\centering
\begin{subfigure}{0.49\columnwidth}
\centering
\includegraphics[width = \columnwidth]{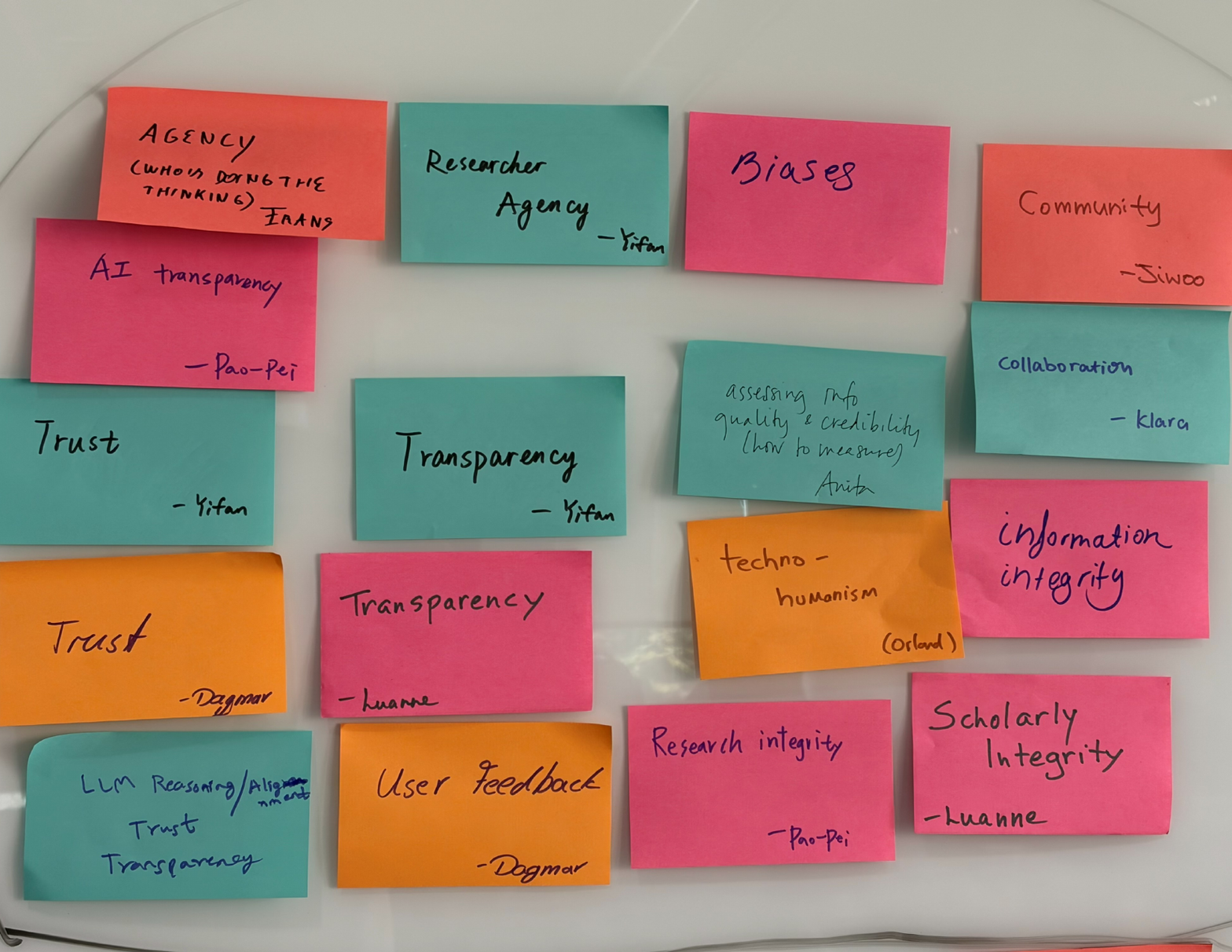}
\caption{Foundations \& Principles}\label{subfig:foundations}
\end{subfigure}
\begin{subfigure}{0.49\columnwidth}
\centering
\includegraphics[width = \columnwidth]{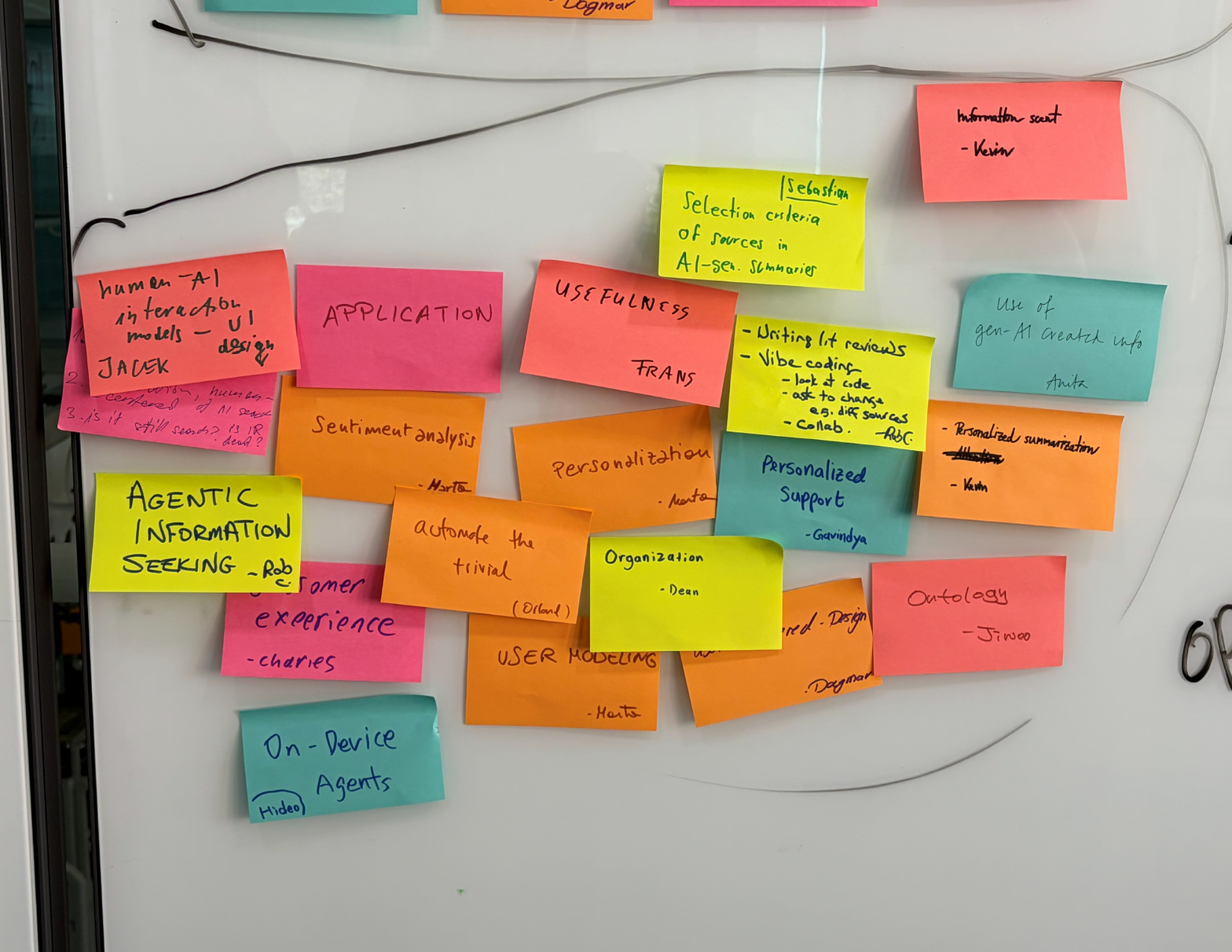}
\caption{Applications \& Opportunities}\label{subfig:applications}
\end{subfigure}
\begin{subfigure}{0.49\columnwidth}
\centering
\includegraphics[width = \columnwidth]{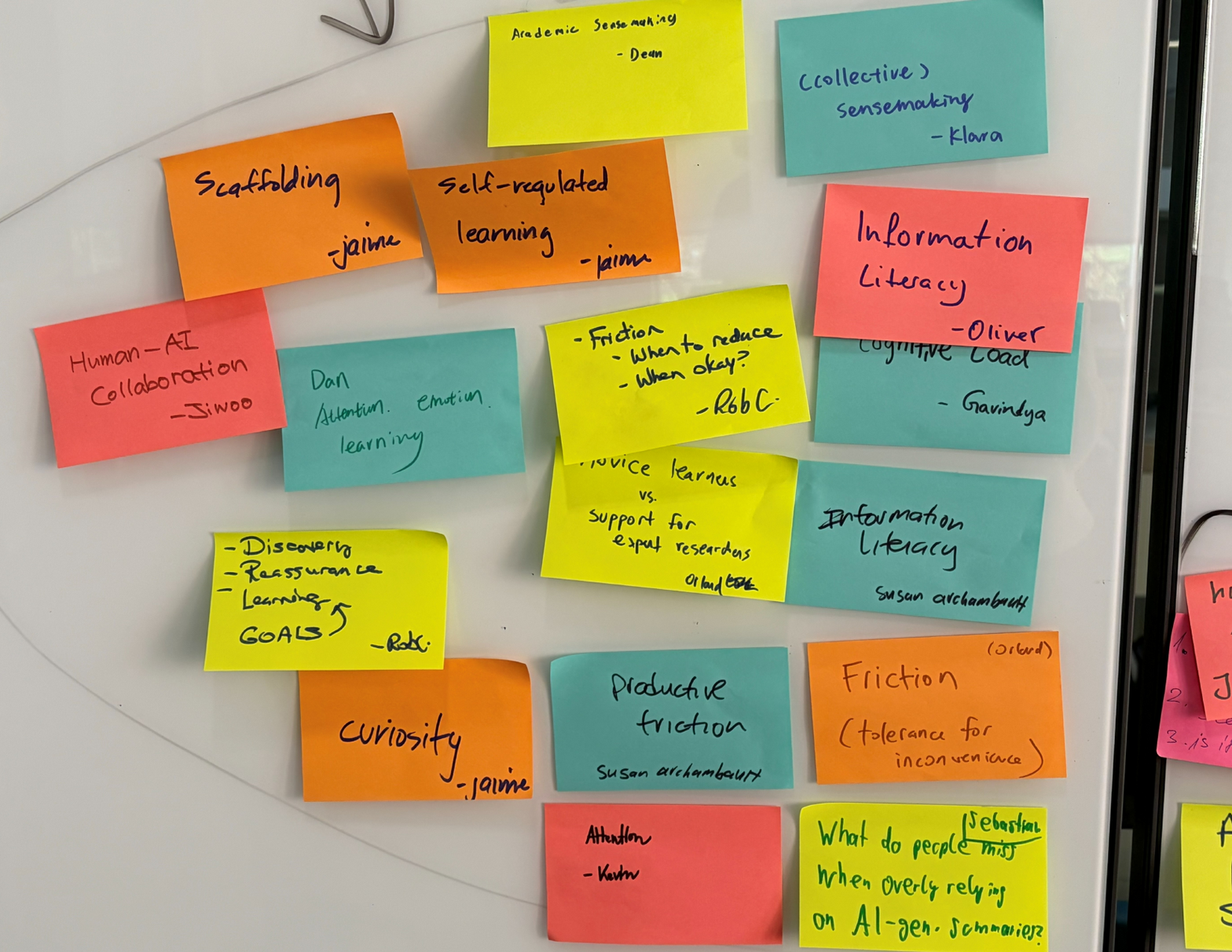}
\caption{Search-as-Learning}\label{subfig:sal}
\end{subfigure}
\caption{Thematic Clusters}
\label{fig:behaviors}
\end{figure}

The first set of roundtable discussions focused on key areas for future work.  To identify these, each workshop participant was asked to write three key areas for future work on sticky notes.  Then, during the first coffee break, the workshop organizers arranged the sticky notes into thematic clusters.  Three thematic clusters were identified: (1) Foundations and Principles (Figure~\ref{subfig:foundations}), (2) Applications \& Opportunities (Figure~\ref{subfig:applications}), and (3) Search-as-Learning (Figure~\ref{subfig:sal}).

The Foundations \& Principles thematic cluster touched upon principles of ``good'' GenAI design, particularly in academic contexts. Popular topics included: user/researcher agency, transparency, trust, user feedback, research/scholarly integrity, and credibility (e.g., lack of bias).

The Applications \& Opportunities thematic cluster touched upon ways in which GenAI technologies may support users in academic contexts.  This cluster also named opportunities for future research.  Applications included: providing personalized support (e.g., personalized summarization), on-device agents, and agentic AI---supporting users with long-term information needs (e.g., writing a literature review) and human-AI collaborations on tasks such as computer programming. Research opportunities included: developing new human-AI interaction models, understanding the uses of AI-generated information in academic contexts, and inferring user satisfaction.

The Search-as-Learning thematic cluster touched upon research related to the use of GenAI tools to support human learning.  The most popular topic was the notion of ``friction''---understanding which (meta)cognitive processes are critical for learning, and developing systems that encourage users to engage with these critical (and possibly uncomfortable) processes \emph{themselves} instead of offloading them to the GenAI.  Other topics included: scaffolding tools to support learning, improving AI literacy, understanding the causes for cognitive load, and understanding the unique needs of domain novices versus experts.

\textbf{Foundations \& Principles:} This roundtable discussion focused on principles of good GenAI design.  GenAI systems should build trust by being more transparent about \emph{process}, \emph{purpose}, and \emph{performance}.  Regarding process, to whatever extent possible, systems should be able to explain how an answer or output was generated (e.g., which sources informed the generated output).  Regarding purpose, to provide the appropriate level of support, systems should elicit information about the user's higher-level purpose.  Regarding performance, systems should be able to convey their level of confidence about the accuracy and completeness of a response.

Several other research opportunities were identified.  First, systems should be able to recognize and highlight possible biases (e.g., confirmation bias) that a user might be demonstrating based on their interactions (e.g., wording of their prompts).  Second, systems should enable users to make connections between different disciplines.  Third, to help educate users, we need to better understand the unique needs of domain novices and experts.  Finally, we need to develop methodologies to measure the extent to which systems exhibit principles of good GenAI design (e.g., transparency regarding process, purpose, and performance).

\textbf{Applications \& Opportunities:} This roundtable discussion enumerated several applications for GenAI systems in academic contexts.  First, systems could provide guidance to users wanting to complete a complex task, which might include suggesting which tools to use for specific subtasks.  Second, systems could proactively monitor the literature on a given topic.  Third, systems could help academic users with literature reviews.  Here, systems could be designed to perform tasks such as: (1) summarizing, comparing, and/or contrasting a list of papers; (2) augmenting a list of papers through forward and backward chaining; and (3) recognizing controversial or debated topics within a list of papers.  Fourth, systems could support users with hypothesis generation, research question formulation, and research gap identification.  Finally, systems could support users with making academic literature accessible to a non-academic audience.

With respect to novel applications, several challenges were identified.  First, given the rapid change of tools, users may not know which tools to use, when, and why.   Second, related to the previous point, research may need to help users develop accurate and complete mental models about a specific tool's purpose and capabilities. Third, for any given application, we need to understand the right balance between actions performed by the GenAI versus the human.  Finally, for any given application, we need to understand the \emph{gulf of execution} and the \emph{gulf of evaluation}.  The gulf of execution asks: What is the gap between what users want systems to do and what existing systems currently support?  The gulf of evaluation asks: How much feedback do users want about the inner workings of a tool? For example, if a specific output is produced, to what extent will users want to understand \emph{how} it was produced?

\textbf{Search-as-learning:} This roundtable discussion focused on opportunities and challenges in developing GenAI systems that encourage and support human learning.  Several opportunities were identified.  First, researchers will need to better understand the differences between good and bad ``friction''.  What are the cognitive processes that should not be offloaded to the GenAI?  How can a system encourage (or nudge) users to engage with those processes?  Second, how can systems adapt to users as they learn?  The idea behind scaffolding is to help learners with tasks they cannot do on their own but to gradually remove the guidance as they become more capable.  Third, systems could help learners by understanding the learner's context.  For example, systems could help students within the context of a specific academic project---by customizing responses to prompts based on the higher-level task.  Finally, systems could tailor responses and elicit information by leveraging what we know about reference librarianship.

Several challenges were also identified.  First and foremost, users may be tempted to offload cognitive processes that are important for learning but may feel uncomfortable.  This relates to the ``principle of least effort''.  Second, basic problems still need to be solved.  For example, systems still hallucinate and users still have difficulty assessing the quality of AI-generated content.  Third, because GenAI systems can respond to highly specific requests and instructions, there may be less opportunity for serendipity and discovery of the ``unknown unknowns''.  Finally, in the long term, AI may compromise people's ability to develop skills such as writing, thinking critically, and thinking creatively, which require prolonged effort.

\subsection{Challenges and Opportunities for the Future}
The second set of roundtable discussions focused on the question: how can we advance human-centered GenAI academic search systems? Three groups of discussants addressed this question from the same set of perspectives: promising methods and approaches, guiding theories and principles for research, and partnerships and community-building activities that would support this work. Each group summarized their discussion collaboratively on a poster that included these three prompts, which were pre-determined by the workshop organizers as components of a forward-looking research agenda. 

\subsubsection{Guiding Theories and Principles}
Workshop participants identified a set of broad principles and commitments to guide research on GenAI academic search systems, as well as a number of theoretical perspectives pertinent to this domain. In addition, they pointed to the need to make visible the goals, values, and ethical commitments guiding research and design work in this space. Recognizing the potential of GenAI tools to increase access to knowledge, academic achievement, and research productivity, participants identified equal access to these tools and their benefits, and the democratization of knowledge, as key commitments. They called for technological humanism over techno-determinism and use of GenAI to augment human abilities rather than replace them. Principles of transparency and explainability were also included, given their importance in enabling user trust and responsible knowledge creation and authorship. 

Regarding the application of theory in this work, participants identified the challenge of operating in a rapidly shifting landscape of norms in academic work, including teaching, learning, research, and publication. They articulated a need for collaborative and interdisciplinary research and theory development between information retrieval and information interaction researchers and those in fields such as scholarly communication, education and social studies of science. At a more granular level, research is needed to assess the relevance of long-standing models, such as the Technology Acceptance Model, to the current context.  Revisiting existing theories and the grounded development of new theories is needed, given the dramatic and potentially transformative impact of GenAI on research and higher education.  

A number of specific theories and approaches central to human information interaction were highlighted as promising for future work in this area, including several that featured in papers presented at the CHIIR 2026 conference. These include: scaffolding and nudging techniques \citep{10.1145/3786304.3787945, 10.1145/3786304.3787866}; friction and effort in learning and engagement; self-determination theory; cognitive load \citep{10.1145/3786304.3787871}; and sensemaking \citep{momeni-hoeber-2026, choi_effects_2025}. 

\subsubsection{Methods and Approaches} Discussions regarding promising methods and approaches fell into several categories. At the broadest level, participants highlighted the need for the CHIIR research community to advocate and ``evangelize''  human-centred approaches to GenAI academic search that fully consider human needs and abilities, in contrast to the dominant technology-focused AI narratives. Suggestions included the publication of white papers and ``manifestos'' to raise awareness and demonstrate the value of small-scale studies of human-system interaction that can produce deep insights, complementing large-scale, automated research agendas. AI literacy and transparency were also associated with advocacy, for example, one group identified the need to develop effective analogies to support more accurate mental models of how these systems work.

A small number of responses reflected areas of work that deserve attention. Suggestions included revisiting slow search ~\citep{10.1145/2528394.2528395} and collaborative search, both human-human and human-AI~\citep{cedenobatistaCollabSearchStudyUserLLM2026} in the context of GenAI academic search. Several suggestions focused on human differences, including cultural, linguistic and neurological diversity, noting the power of GenAI search systems to support diversity through enhanced tools for accessibility, deep personalization and contextual understanding. Participants also identified the need for user studies to understand how GenAI search systems are being used by diverse groups within academia.  

The largest number of responses identified promising research methods for human-centred research. Participants emphasized the need for authentic, participant-engaged research practices that can identify impacts beyond efficiency and effectiveness. Broader use of methods from other disciplines was also identified as an important strategy. Specific suggestions of methods included naturalistic, observational and longitudinal user studies, participatory and value-sensitive design, community-engaged research, and human-in-the-loop (or AI-in-the-loop) approaches. The need for evaluation of short and long term outcomes of GenAI academic search systems on users' affect, behaviour and cognition, on task outcomes, and on scholarship as a whole was emphasized. Participants proposed use of counterfactual study designs, control groups, and pre- and post-task assessments to determine impacts. The possible use of GenAI to design improved user studies was also identified as an area for future work.

\subsubsection{Partnerships}
The third prompt for discussion asked participants to consider the types of partnerships and joint actions needed to advance this work.  Given the high level of interest and engagement among workshop participants, future workshops on this topic at CHIIR and/or other venues are warranted. Partnerships between academic researchers and the information, publishing and technology companies that develop GenAI academic search tools are needed, both to enable broader access to real-world usage data and to support evidence-based alignment of these tools with the needs and values of academic users. This may be accomplished through research internships and more formal research collaborations and joint projects. As institutions on the front lines of GenAI academic search system adoption and use, university libraries also offer enormous potential as research partners for studying adoption, search literacy and training, and conducting joint system evaluations. Some workshop participants are already engaged in such collaborations or conducting this research directly within academic libraries and we can build on these foundations in future workshops. In general, participants expressed that cross-sector, community-engaged, interdisciplinary and international collaborations are necessary to move this research forward to allow for comprehensive and inclusive, multilingual and multicultural approaches.

\section{Continuing Activities}


GenAI academic search systems constitute a new class of information tools with capabilities well beyond traditional bibliographic databases and academic search engines. This is a disruptive technology within the space of academic research, teaching and learning, which offers enormous opportunities and raises substantial questions and concerns. The GAI\&AS workshop brought together researchers in human information interaction and retrieval for an initial exploration of this topic to help build a future research agenda. The workshop demonstrated substantial interest and a wide range of research initiatives underway.  Building on these first steps, we plan to continue to expand and strengthen the community. Plans for a second GAI\&AS workshop at a future CHIIR meeting (e.g., Berlin, 2027) are underway. Drawing upon ideas from the current workshop, we will extend invitations to participate in the workshop to academic librarians and researchers and developers within the academic search and publishing industries. We will seek opportunities for a special issue of a journal on this topic. For example, there will be an Academic Search and Learning with GenAI Track at the 2026 Information Processing \& Management Conference (IP\&MC2026).

\appendixauthorothers

\textbf{First author tier:}
\begin{itemize}
    \item Yifan Liu, University of British Columbia, Canada
    \item Jaime Arguello, University of North Carolina at Chapel Hill, USA.
    \item Orland Hoeber, University of Regina, Canada.
    \item Chang Liu, Peking University, China.
    \item Soo Young Rieh, University of Texas at Austin, USA.
    \item Luanne Sinnamon, University of British Columbia, Canada
\end{itemize}

\textbf{Second author tier:}
\begin{itemize}
	\item Dean Alvarez (University of Illinois, USA).
    \item Susan Archambault (Loyola Marymount University, USA).
    \item Rob Capra (University of North Carolina at Chapel Hill, USA).
    \item Henson Chen (University of Washington, USA).
    \item Charles Costa (Lexora Labs, USA).
    \item Anita Crescenzi (University of North Carolina at Chapel Hill, USA).
    \item Zhitong (Klara) Guan (University of Texas at Austin, USA).
    \item Jacek Gwizdka (University of Texas at Austin, USA).
    \item Pao-Pei Huang (University of North Carolina at Chapel Hill, USA).
    \item Gavindya Jayawardena (University of Texas at Austin, USA).
    \item Ghazal Kalhor (University of Tehran, Iran)
    \item Dagmar Kern (GESIS Leibniz Institute for the Social Science, Germany).
    \item Oliver Koop (Hamburg University of Applied Sciences, Germany).
    \item Alice Li (University of British Columbia, Canada).
    \item Afra Mashhadi (University of Washington, USA).
    \item Gaohui Meng (Peking University, China).
    \item Marta Micheli (University of Turin, Italy).
    \item Anil B. Murthy (Arizona State University, USA).
    \item Kevin Schott (GESIS Leibniz Institute for the Social Science, Germany).
    \item Sebastian Schultheiß (Hamburg University of Applied Sciences, Germany).
    \item Jiwoo Seo (Florida State University, USA).
    \item Phaneendra Sivangula (Rutgers University, USA).
    \item Frans van der Sluis (University of Copenhagen, Denmark).
    \item Xiaoxuan Song (Nanjing Agricultural University, China).
    \item Silang Wang (University of Washington, USA).
    \item Dan Zhang (University of Texas at Austin, USA).
\end{itemize}

\bibliography{
sigirforum, lt-hoeber, AISS, lt-arguello, lt-archembault, lt-schott, lt-liu-meng-song, lt-seo, chiir-submission-bib}
\end{document}